\documentclass[12pt]{iopart}
\usepackage[utf8]{inputenc}
\usepackage[square,numbers]{natbib}
\usepackage{graphicx}
\usepackage{subfigure}
\usepackage{siunitx}
\usepackage{placeins}
\usepackage{color}
\usepackage{standalone}
\usepackage{dcolumn}
\usepackage{bm}
\usepackage{microtype}
\usepackage{etoolbox}
\usepackage{amssymb}
\usepackage{accents}
\usepackage[normalem]{ulem}
\usepackage[dvipsnames]{xcolor}
\usepackage[colorlinks,urlcolor=NavyBlue,citecolor=NavyBlue,linkcolor=NavyBlue,pdfusetitle]{hyperref}
\usepackage[all]{hypcap}
\usepackage{enumerate}
\usepackage{enumitem}
\usepackage{gensymb}
\usepackage{comment}

\newcommand{\jade}[1]{{\color{black}{#1}}}

\begin{document}

\title[Generating transient noise]{Generating transient noise artifacts in gravitational-wave detector data with generative adversarial networks}
\author{Jade Powell$^{1,2}$, Ling Sun$^{1,3}$, Katinka Gereb$^{4}$, Paul D. Lasky$^{1,5}$, Markus Dollmann$^{4}$} 

\address{$^{1}$OzGrav: The ARC Centre of Excellence for Gravitational-wave Discovery, Australia}
\address{$^{2}$Centre for Astrophysics and Supercomputing, Swinburne University of Technology, Hawthorn, VIC 3122, Australia}
\address{$^{3}$OzGrav-ANU, Centre for Gravitational Astrophysics, College of Science, The Australian National University, ACT 2601, Australia}
\address{$^{4}$Eliiza, Level 2/452 Flinders St, Melbourne VIC 3000}
\address{$^{5}$School of Physics and Astronomy, Monash University, Clayton, VIC 3800 Australia}
\ead{jade.powell@ligo.org}
\vspace{10pt}
\begin{indented}
\item[]\today
\end{indented}


\begin{abstract}

Transient noise glitches in gravitational-wave detector data limit the sensitivity of searches and contaminate detected signals. In this Paper, we show how glitches can be simulated using generative adversarial networks (GANs). We produce hundreds of synthetic images for the 22 most common types of glitches seen in the LIGO, KAGRA, and Virgo detectors. We show how our GAN-generated images can easily be converted to time series, which would allow us to use GAN-generated glitches in simulations and mock data challenges to improve the robustness of gravitational-wave searches and parameter-estimation algorithms.
We perform a neural network classification to show that our artificial glitches are an excellent match for real glitches, with an average classification accuracy across all 22 glitch types of 99.0\%. 
    
\end{abstract}

\section{Introduction}

The Advanced LIGO \citep{2015CQGra..32g4001L}, Advanced Virgo \citep{2015CQGra..32b4001A}, and KAGRA \citep{2020arXiv200505574A} gravitational-wave detectors are currently offline, being upgraded and prepared for their fourth observing run (O4). The previous three observing runs have resulted in the successful detection of a population of gravitational-wave signals from merging compact binaries made up of neutron stars and black holes \citep{2019PhRvX...9c1040A, 2021PhRvX..11b1053A, 2021arXiv211103606T}.
The number of detectable gravitational-wave events will significantly increase during O4, which will result in the analysis of a larger number of gravitational-wave signals being negatively impacted by transient noise artifacts in the detectors, known as glitches \cite{2021CQGra..38m5014D, 2022arXiv220501555A, 2021PTEP.2021eA102A}.

Glitches limit the sensitivity of transient gravitational-wave searches. Glitches that occur concurrently in multiple detectors can result in false alarms for searches. They can also overlap with gravitational-wave signals, making it difficult to measure the astrophysical parameters of the source. For example, the first gravitational-wave detection of a binary neutron star merger GW170817~\citep{2017PhRvL.119p1101A} had a large glitch in the LIGO Livingston detector data $\sim1$\,s before the merger time. A further sixteen signals in the third observing run were contaminated by glitches \citep{2021PhRvX..11b1053A, 2021arXiv211103606T}. Before parameter estimation of these signals can begin, glitches are often reconstructed and removed using a tool called BayesWave \citep{2015CQGra..32m5012C, 2021PhRvD.103d4006C}. Although this method has been very successful at glitch removal, it requires a lot of time and computing power that will make glitch subtraction difficult when the number of gravitational-wave signals significantly increases. Therefore, efforts to reduce the number of glitches that occur in gravitational-wave data are important for future gravitational-wave searches and parameter estimation. 

Glitch classification is one of the methods currently used for the reduction of glitches in gravitational-wave detector data. Glitch classification algorithms sort glitches into different types according to their morphology in spectrograms. Knowing the times that glitches of the same type occur may allow us to find the origin of the glitches and eliminate them. Glitch classification algorithms typically use machine learning techniques \cite{2020arXiv200503745C}. The first glitch classification attempts used simple Gaussian mixture models to classify the glitches into different types \citep{2015CQGra..32u5012P, 2017CQGra..34c4002P}. This was further improved by the use of neural networks \citep{2018CQGra..35i5016R, 2017arXiv171107468G, 2021arXiv211110053S}, however these require large pre-labelled sets of glitches for training the networks. The labelled glitch training sets for neural networks are currently provided by the citizen~science project Gravity Spy~\citep{2017CQGra..34f4003Z, BAHAADINI2018172, 2021CQGra..38s5016S}, where users classify spectrograms of glitches by visual inspection of the images. 

Artificial but realistic glitches are required to produce mock gravitational-wave data for testing the robustness of searches and parameter estimation tools in the presence of glitches. In Refs.~\cite{2022arXiv220306494L,2022arXiv220509204L}, generative adversarial networks (GANs) \cite{goodfellow2014generative} are used to produce fake time-series waveforms for the most common type of detector glitch, known as blips. GANs are a type of neural network that can be used for creating new, synthetic data that mimic their input. In \cite{2022arXiv220306494L, 2022arXiv220509204L} the authors produce the time-series waveforms for the blips using BayesWave. They then feed their blip waveforms into the GAN to produce fake blip glitches that can be used to test search and parameter estimation pipelines. However, there are a large number ($\sim25$) of different types of glitches that occur in gravitational-wave data. The blip glitch is the most common type, but a real gravitational-wave signal could overlap with any of the other types of glitches. In addition, using BayesWave is a computationally expensive method for producing a training set for a GAN. Gravitational-wave searches using neutral networks also usually require images of glitches for training instead of time-series waveforms. 

In this study, we expand on the previous work by using GANs to produce artificial images for 22 different types of glitches. For training, we directly use the previously classified glitches from Gravity Spy, so no significant further computational resources are required for pre-processing the images before training. We run a neural network classification on the artificial glitches to measure the performance of the GAN. \jade{We show how our GAN-generated images can easily be converted into time series for the use in simulations and mock data challenges to improve gravitational-wave search and parameter-estimation codes.}

This Paper is structured as follows: In Section \ref{sec:method} we give an overview of the neutral networks and GANs used in this study, and describe the data and training. The results of the GAN-generated glitch images are presented in Section \ref{sec:results}. \jade{ Applications of our GAN-generated glitches are discussed in Section \ref{sec:app}. } A discussion and conclusion are given in Section \ref{sec:concl}.

\section{Methodology}
\label{sec:method}

We split our methodology into three parts. In Section \ref{sec:cnn}, we describe Convolutional Neural Networks (CNN) \cite{Lawrence1997, 2015arXiv151108458O}, which are a component of the GAN algorithm, and are also used to measure how realistic the GAN-generated images are. In Section~\ref{sec:gans}, we briefly describe the GAN algorithm that we use to generate various types of fake glitches by training on the real labelled glitches from Gravity Spy. We describe details about the data sets, setup of the network architecture, and data pre-processing in Section~\ref{sec:data}.

\subsection{Convolutional Neural Networks}
\label{sec:cnn}

\begin{figure}[!t]
    \includegraphics[width=16.0cm]{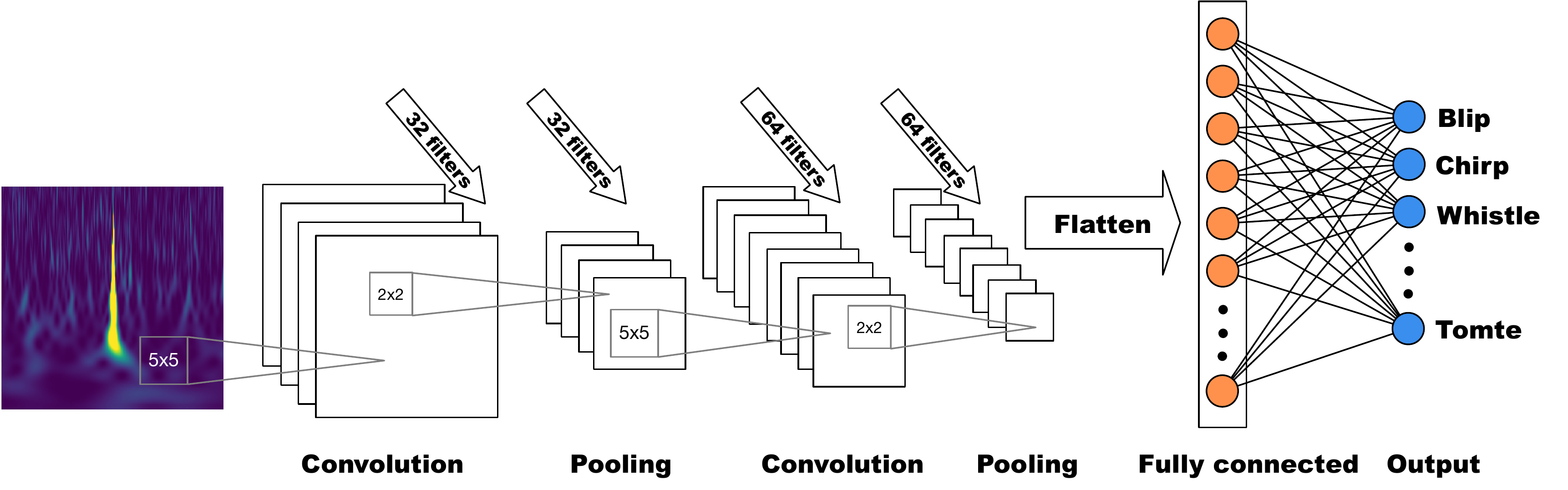}
    \caption{A schematic of our CNN architecture used to classify glitches in gravitational-wave detectors. We use two convolutional layers, two max pooling layers, and one fully connected layer including the flattened input and output. After each pooling layer and in the fully connected layer, we use a dropout rate of 50\%. The squares marked by ``$5 \times 5$'' and ``$2 \times 2$'' indicate the convolutional and pooling kernel sizes, respectively.}
    \label{fig:cnn}
\end{figure}

CNNs have been widely used for solving classification problems in gravitational-wave astronomy \cite[e.g.,][]{2018CQGra..35i5016R, 2017arXiv171107468G, 2017CQGra..34f4003Z, 2022arXiv220207158Z}.
Our CNN architecture is shown in Figure~\ref{fig:cnn}. 
First, convolution kernels are applied to the data to extract features present in the images. 
Max pooling is then carried out by extracting the maximum value in the applied feature maps to preserve meaningful information, and downsampling the rest.
We use a kernel size of ``$5 \times 5$'' and ``$2 \times 2$'' pixels for convolution and max pooling, respectively, indicated by the small grey squares in Figure~\ref{fig:cnn}. A dropout rate of 50\% is used during training after each pooling layer, which helps to prevent overfitting.
We use 32 feature maps (or filters) in the first layers of convolution and pooling, and use 64 of them in the second layers. 
At the end, the data is flattened into one array and forward propagated through the neurons of the network via a fully connected layer.
Again, a 50\% dropout rate is used in the fully connected layer.
Weights are applied to the data points in each neuron, and the weighted sum of the input signal is then passed through the final layer by the neurons’ activation function. The output is where predictions are produced, and a loss function is used to compute the distance between the predicted and expected outcomes. This, together with the learning rate, dictates how much the weights are to be updated in the next learning epoch. This process is repeated for several learning epochs as the neurons ``learn'' what features are important and need to be passed on to produce the desired predictions.

\subsection{Generative Adversarial Networks}
\label{sec:gans}

GANs can create new, synthetic data that mimic the real input by the application of multiple CNNs. GANs have been widely used for various applications, e.g., text and image translation \cite{2018arXiv181204948K}, time series generation \cite{2021arXiv210711098B}, and producing artificial images of Galaxies \cite{2017MNRAS.467L.110S}. 

\begin{figure}[!t]
    \includegraphics[width=16.0cm]{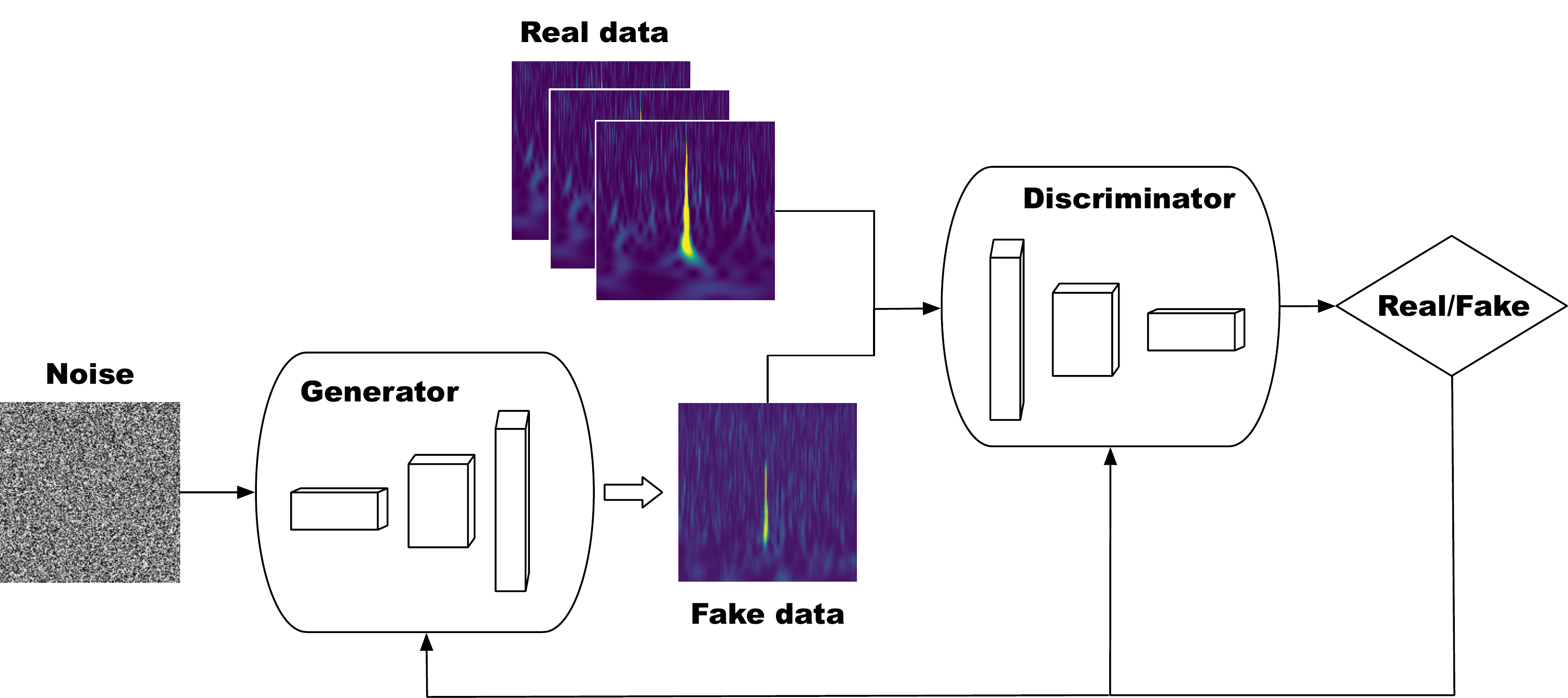}
    \caption{A schematic of our GAN architecture that is used to produce artificial images of detector noise glitches. The generator and discriminator consist of multi-layer neural networks. The generator produces the fake glitches. The discriminator tries to distinguish the fake from real glitches.}
    \label{fig:gans}
\end{figure}

GANs use two CNNs, a generator and a discriminator, where the former is responsible for generating fake images from the input random noise, and the latter is used for discriminating the real and generated data, as shown in Figure~\ref{fig:gans}. 
The generator progressively learns the features of the real images and generates fake images mimicking the real ones through backpropagation.
In this study, we use TorchGAN~\cite{Pal2021}, an implementation of the Deep Convolutional Generative Adversarial Network (DCGAN) architecture. 

\subsection{Data and training}
\label{sec:data}

\subsubsection{Gravitational-wave data}
\label{sec:gravityspy}

We use the data set provided by Gravity Spy \cite{coughlin_scott_2018_1476551} that was used to train their CNN after being classified by citizen scientists. The data set contains 7966 time-frequency images of glitches collected in the first and second observing runs of Advanced LIGO. The time-frequency spectrogram images are produced by performing a Q-transform \cite{2004CQGra..21S1809C}, which works by using a sine-Gaussian template decomposition.

There are 22 different glitch types present in the data. Glitches are only included if they occur when the detector is in observing mode. They all have a signal-to-noise ratio larger than 7.5, and a peak frequency in the most sensitive frequency band of the detectors, between 10\,Hz and 2048\,Hz. Some of the glitches have a known physical cause, and the origin of the other glitches is still unknown. The number of glitches in each type differs significantly as the glitches that occur in the data change as the environment and detector conditions at the site alter during the observing run. 

Each glitch is defined by four different views (with resolutions of $140 \times 170$ pixels) over four durations: 0.5\,s, 1.0\,s, 2.0\,s and 4.0\,s. 
Empirical results show that using the 2.0\,s duration images produces the most accurate classification.
Therefore, for the rest of the paper, we use only the 2.0\,s images.

\subsubsection{CNN}
\label{sec:cnn_details}

To test the accuracy of our GAN-generated glitches, we perform glitch classification using a CNN with the same model architecture as the Gravity Spy paper \cite{BAHAADINI2018172}.
A summary of the model architecture is given in Table \ref{tab:cnn_table}.
The models are allowed to train for 100 epochs, with a checkpoint at the end of every training epoch to update the weights of the best model. Early stopping is applied to the model with a patience of 20, i.e., if the model performance does not improve over 20 epochs, training stops.

\begin{table}[!t]
	\caption{\label{tab:cnn_table} A summary of the CNN model architecture. The numbers in parentheses in the second column indicate the number of kernels. In the last row, the output shape number, 22, corresponds to the 22 types of glitches.}
	\begin{indented}
	\item[]\begin{tabular}{@{}ll}
	\br
	Layer & Output shape  \\
	\mr
	Conv2D & $140 \times 170$ (32)    \\
	MaxPooling2D & $70 \times 85$ (32)    \\
	Dropout & $70 \times 85$ (32)  \\
	Conv2D & $70 \times 85$ (64)    \\
	MaxPooling2D & $35 \times 42$ (64)    \\
	Dropout & $35 \times 42$ (64)  \\
	Flatten & 94080  \\
	Dense & 256  \\
	Dropout & 256  \\
	Dense & 22  \\
	\br
	\end{tabular}
	\end{indented}
\end{table}

A few pre-processing steps are taken to prepare the training data sets. We first filter out a subset of featureless images (mainly in the ``extremely loud'' and ``paired doves'' categories), as well as a small number of misclassified, or visually undetectable glitches.
Next, there exists a significant class imbalance among different glitch types. The smallest class is the ``paired doves'' with only 18 examples, and the largest one is ``blip'' glitches with over 1750 examples. Deep learning algorithms are not expected to provide good predictive accuracy in the case of small sample sizes. 
Thus for glitch classification, we conduct a partial oversampling/undersampling, by applying some oversampling to small classes and undersampling to larger ones, instead of setting the exact same sample size for all classes.
The oversampling/undersampling sizes are determined empirically. Small glitch classes are oversampled to a minimum of 250 sample size. This oversampling is achieved by duplicating the samples. Even though it does not add new information, data oversampling is an effective and efficient approach for achieving better prediction accuracy for small sample sizes.
The maximum sample size for categories with more than 250 samples are upsampled to 600, (close to the actual upper limit among different glitch types, except for the largest category ``blip''). An exception to the upsampling rule is the blip category, which is the largest class with 1750 samples. Thus, for blip glitches, we allow a larger sample size of 1000. 
After oversampling, the performance of some of the smaller categories such as whistle, no glitch and low frequency lines was still lagging behind, therefore a maximum sample size of 600 was allowed for these classes.
Further details are provided in \ref{appendix:balancing} for the number of examples in each class before and after the oversampling/undersampling.

\subsubsection{GAN}
\label{sec:gan_details}

For all glitch classes, an oversampled size of 5000 is used. GANs require a large amount of training data, and our results were significantly improved with this level of oversampling. TorchGAN only accepts square input images with sizes that are a power of two. Consequently, the $140\times170$ dimensional arrays are converted to a size of $256 \times 256$ by adding zeros as padding. 

We create a separate GAN model for each glitch type. The models are trained for 500 epochs using a batch size of 100. TorchGAN creates five different models that are saved at consecutive iterations as training progresses. 
All other glitch types share the same GAN model, except for the ``violin mode,'' which uses a different GAN model to achieve better results in regards to the glitch morphology. 

In most cases, a Wasserstein loss function is used. 
It is observed in some categories, e.g., ``1080 lines'', that the generated images can look similar to each other without much variety in their own category. 
This effect is due to mode collapse, where the gradients of the discriminator fail to make the generator adequately learn the true distribution. 
In the case of ``1080 lines'', ``1400 ripples'', ``whistle'', and ``tomte'', more varied outputs are achieved by using a different loss function: the least square loss function.

By default, a hyperbolic tangent activation function is used in the last convolutional layer of TorchGAN's generator function, which produces output values in the range of $(-1,1)$. 
However, we use a sigmoid activation function to achieve output values in a similar range as those in the input images, i.e., $(0,1)$.

\section{Results}
\label{sec:results}

\begin{figure}
    \centering
    \includegraphics[width=15.0cm]{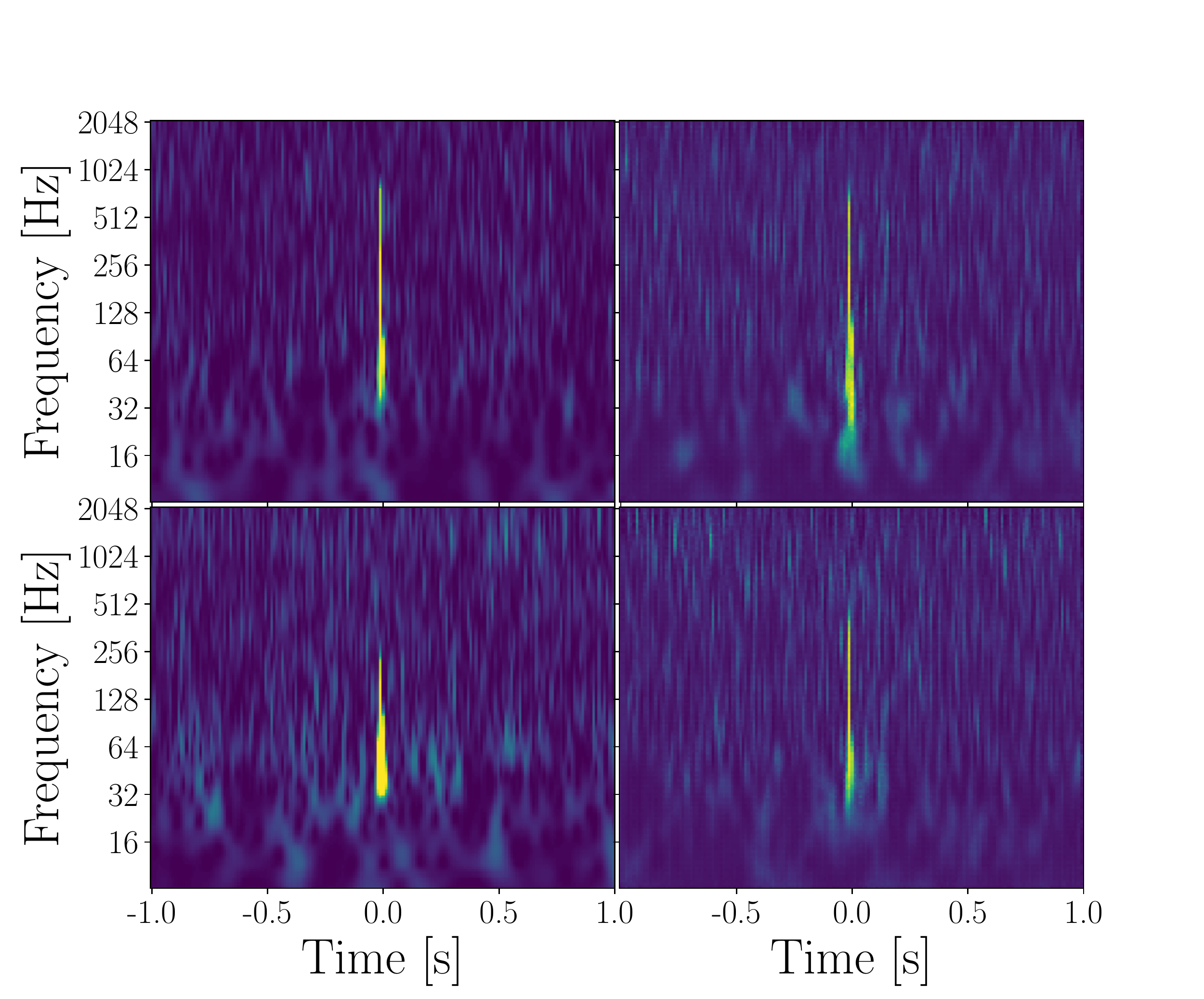}
    \caption{Blip glitches. Two fake blip glitches generated with the GAN (right column) and two real glitches (left column). The colourscale is in arbitrary units.}
    \label{fig:fake_blips}
\end{figure}

The GAN models are used to generate 100 fake images for each of the 22 types of glitches. Examples of the comparison between real and GAN-generated glitches are shown in Figure \ref{fig:fake_blips} for blip glitches and Figure \ref{fig:fake_koi} for Koi fish glitches. Further examples for other glitch types are shown in \ref{appendix:all_glitches}. We find that all of our simulated glitches are a good representation of the original real input glitches. The GAN-generated glitches show the main features of the real ones and cannot be distinguished by eye from the training set. We also see variation within the different glitch classes, demonstrating that our GAN models do not suffer from mode collapse, which would cause every generated fake glitch in a class to look identical. 

\begin{figure}
    \centering
    \includegraphics[width=15.0cm]{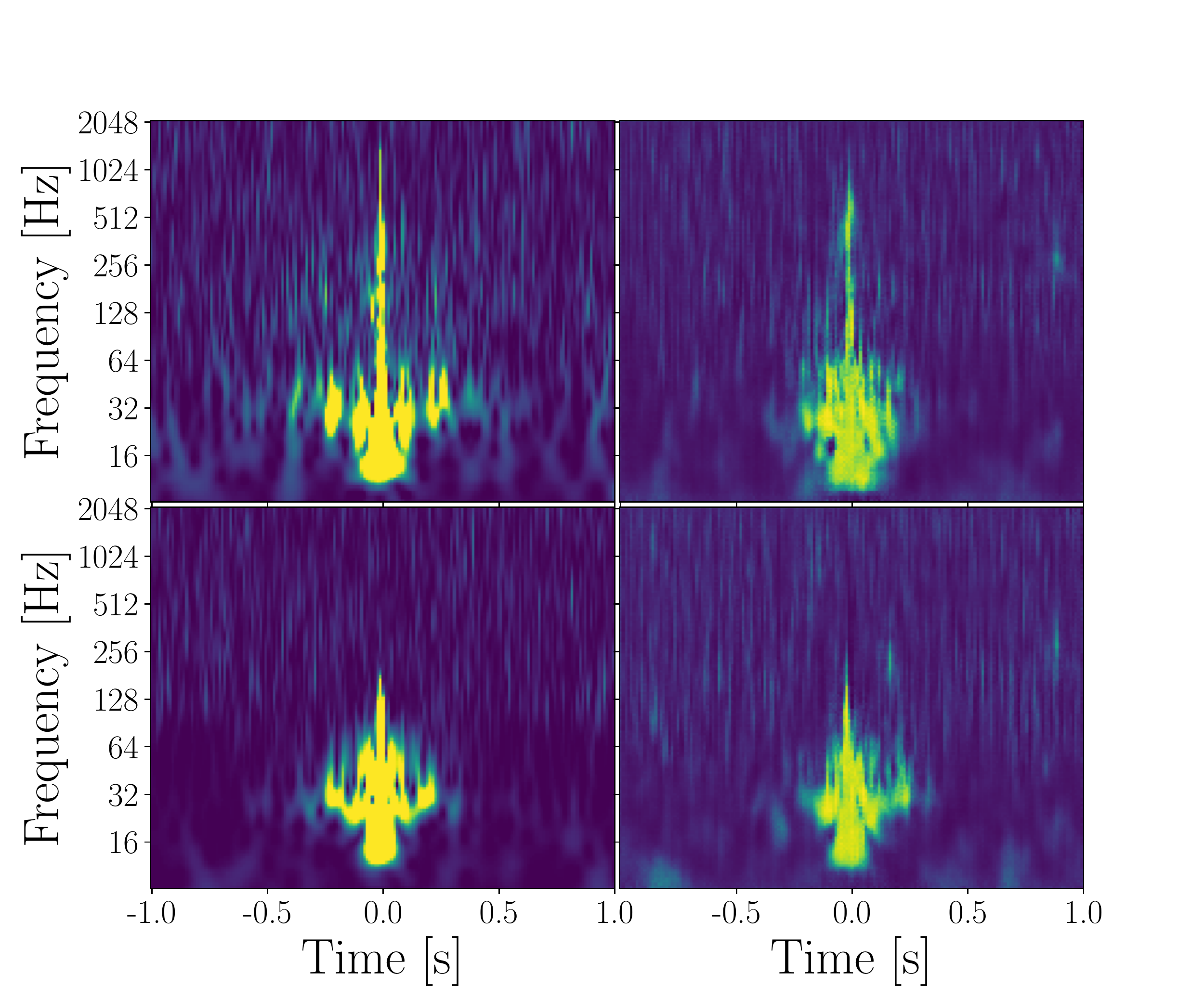}
    \caption{Koi Fish glitches. Two fake Koi Fish glitches generated with the GAN (right column) and two real glitches (left column). The colourscale is in arbitrary units.}
    \label{fig:fake_koi}
\end{figure}

To test if the fake glitches are a good representation of the real input glitches, the compiled dataset of 2200 fake glitches are used as a test dataset for a CNN model, which is then applied to predict the glitch categories of the GAN-generated glitches. If the fake images are realistic enough, the model should be able to predict their glitch category at a high accuracy. We use the same set up for our CNN glitch classifier as is used in the Gravity Spy glitch classification project, as described in Section \ref{sec:cnn_details}. 

\begin{figure}
    \includegraphics[width=16.5cm]{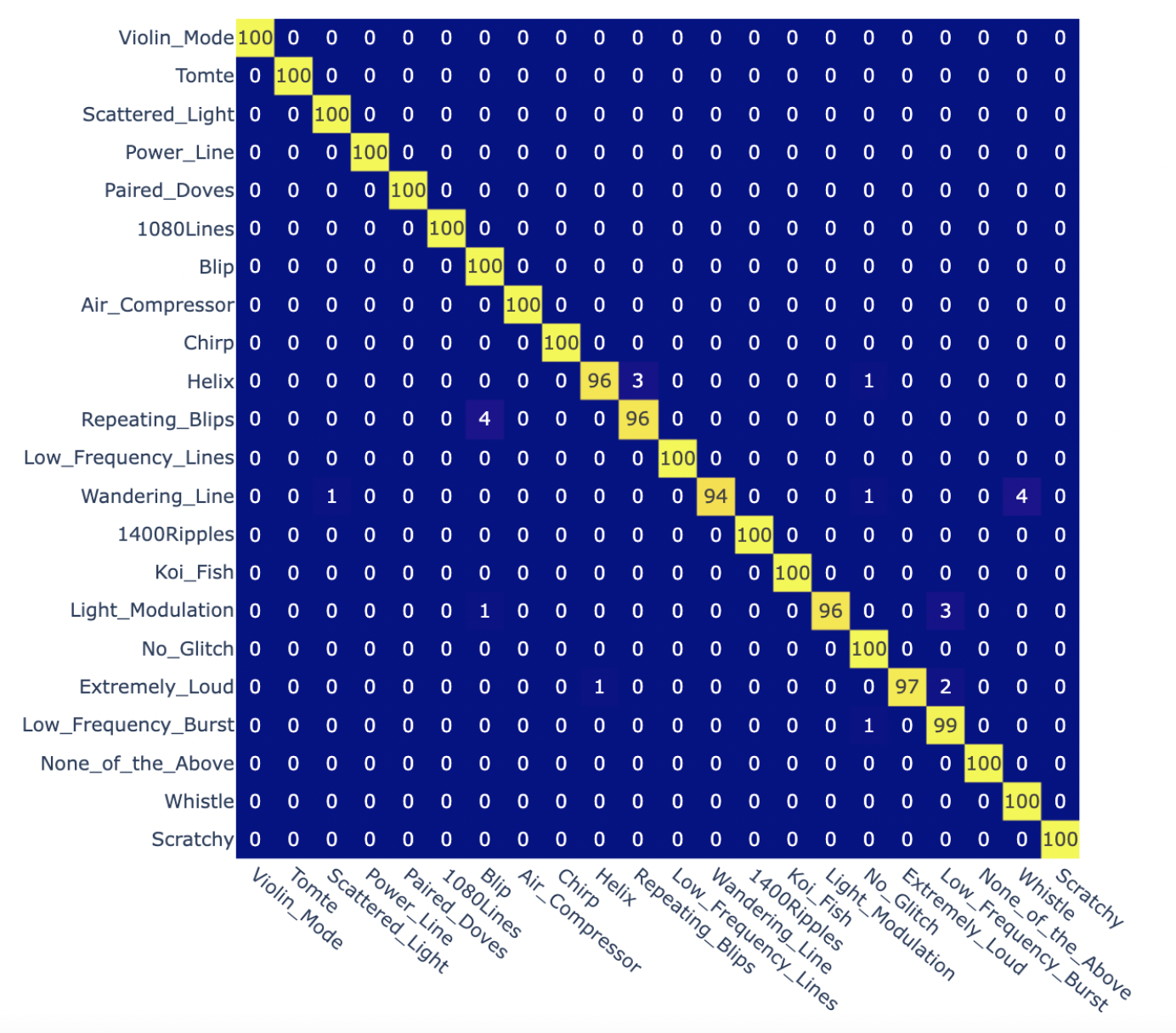}
    \caption{Confusion matrix showing the CNN classification of the fake glitches generated with the GAN. All GAN-generated glitches are classified with a high accuracy, indicating that they are a good representation of the original training glitches. }
    \label{fig:classification}
\end{figure}

In Figure \ref{fig:classification}, we show the CNN prediction accuracy for all the generated fake glitches produced by the GAN. 
The CNN model is able to predict the different glitch categories at a high accuracy on the fake data generated by the GAN. 

\section{Applications}
\label{sec:app}

\begin{figure}
    \includegraphics[width=16cm]{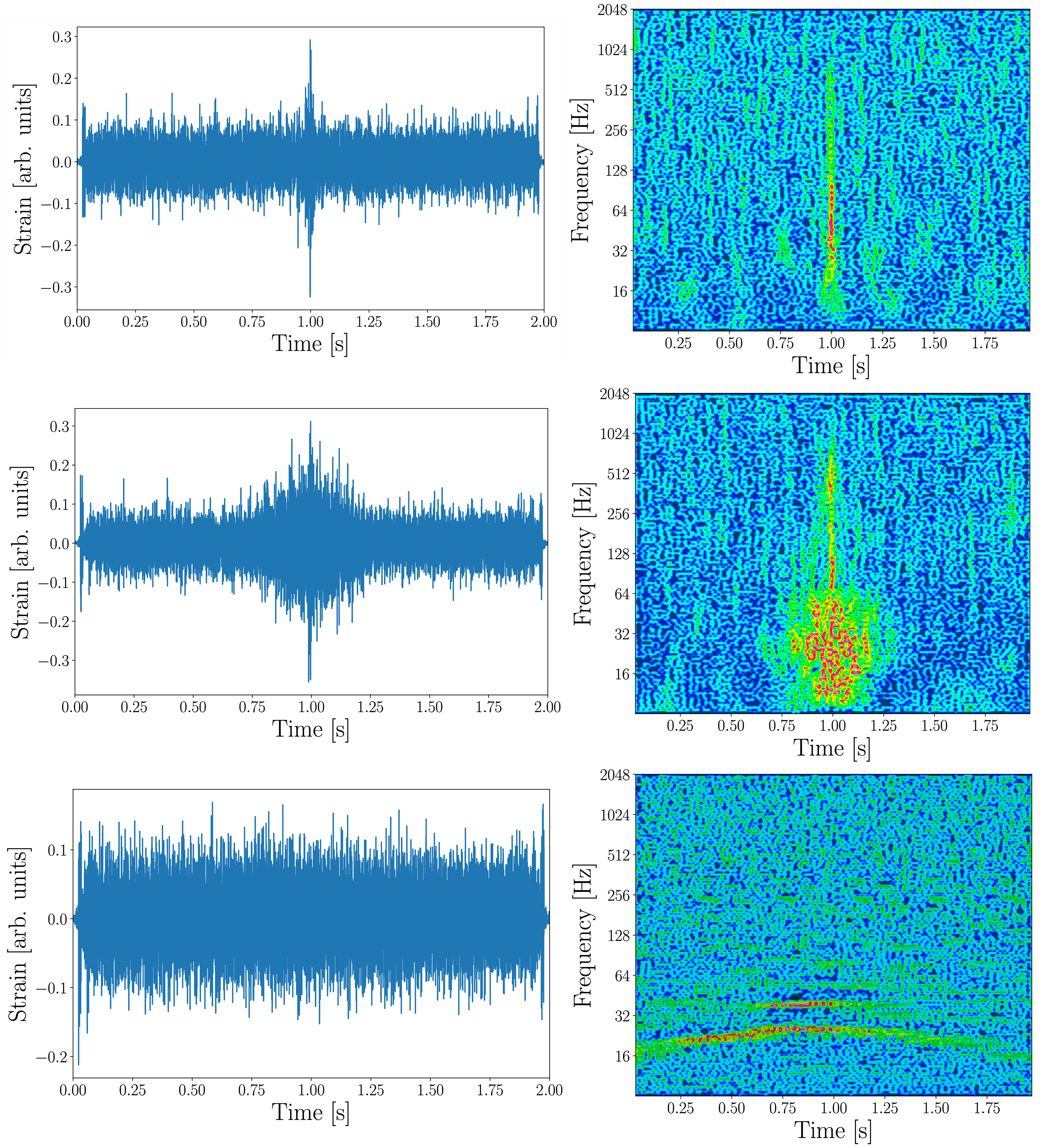}
    \caption{\jade{Time series data converted from GAN-generated glitch images (left), and the reconstructed spectrograms of the time series (right).  
    From top to bottom: blip glitch, Koi fish, and scattered light.}}
    \label{fig:time}
\end{figure}

\jade{
In this Section, we describe some of the potential uses of artificial glitches generated by GANs. 

The first potential application is to directly use the GAN images to improve glitch classification algorithms like GravitySpy. There is a significant imbalance in the number of glitches in each category. Using GANs to up-sample glitch categories before classification may produce better results than the method we use to up-sample in \ref{appendix:balancing}. An example of this has also been shown in previous work \cite{2022MNRAS.515.4606Y}. 

Image-based searches are also becoming increasingly common in gravitational-wave data analysis. Some examples involve searches for gravitational waves from compact binaries \cite{2022arXiv220911146S} and core-collapse supernovae \cite{2018PhRvD..98l2002A}. Our GAN-generated glitch images could be used in the training sets of those image-based gravitational-wave searches to improve their robustness against detector noise.  

Standard gravitational-wave searches and parameter estimation techniques typically use time-series data. Time-series glitches can be produced directly from our GAN-generated images. The most common method for doing this is described in Ref.~\cite{1164317}. We show some examples of the time series in Figure \ref{fig:time}, produced from a GAN-generated blip glitch (top panel), Koi fish (middle panel), and scattered-light glitch (bottom panel). Converting the images into time-series glitches takes less than a second. 
To produce the time-series glitches in the left-hand column of Figure \ref{fig:time}, we use the \texttt{InverseSpectrogram} function in \texttt{Mathematica}. We then convert the time series (left) back to spectrograms (right) for a sanity check and comparison with the original images. The spectrograms clearly have the same glitch features as the GAN-generated images.

Although the recovered spectrograms are not identical to the GAN-generated ones due to some missing information in the phase conversion, the main structures are sufficiently preserved for simulation purposes.
This process can enable time-series glitches to be injected into gravitational-wave data to verify gravitational-wave search algorithms when a glitch is present in the data. Time-series glitches could also be injected on top of, or close to, signals to test parameter-estimation codes. 

}

\section{Conclusions}
\label{sec:concl}

Noise glitches in gravitational-wave detectors limit the sensitivity of gravitational-wave searches and contaminate real signals. Efforts to remove them from the data are computationally expensive, therefore eliminating their origin is important for fast and accurate gravitational-wave search and parameter estimation for astrophysical source properties. 

For gravitational-wave data, previous studies have applied GANs to generate time series of fake gravitational-wave burst signals and fake blip glitches. However, there are many other types of glitches that could contaminate signals, and some algorithms, for example machine learning searches, require images for training. Therefore, in this study, we apply a GAN to a pre-labelled set of 22 different types of glitches. We generate images of fake glitches covering a larger variety of glitches than in previous studies, \jade{and show how those images can easily be converted into time series.}

We use TorchGAN architecture GAN. 
Our model is trained on the pre-labelled data set provided by Gravity Spy from the Advanced LIGO detectors first and second observing runs. We produce 100 fake glitches for each of the 22 different classes of glitches. All of our GAN-generated images are a good match for the original images fed into the GAN. To test this, we perform a neural network classification of the GAN-generated images. Almost all of the fake glitches are classified into the correct glitch class, demonstrating that the GAN-generated glitches are a good match for the original true glitches. 

There are many areas of gravitational-wave science in which our GAN-generated glitches could be applied. The fake glitches could be used to train machine learning searches for transient gravitational waves, such as compact binary mergers or gravitational-wave bursts. Previous machine learning training studies have had to either produce an analytical model for certain glitch types, which does not exist for every type of glitch, or train using the Gravity Spy glitches, which only contain a small number of glitches for some types. Generating fake glitches with a GAN can improve glitch classification by solving the issue of unbalanced classes. \jade{We note that currently, if new types of glitches are identified they will initially be classified as ``None of the Above''---e.g., see Figure~\ref{fig:all_glitches}. Once enough of these glitches are recognised by eye, the GAN can be adapted to create artificial glitches in the way outlined in this work.} 

The GAN-generated glitch images can also be \jade{converted into time series and} used to produce data sets for more traditional gravitational-wave searches. Glitches degrade the sensitivity of gravitational-wave searches, so using fake data to test robustness of these search pipelines is essential before they begin to analyze real gravitational-wave data in low latency. Fake glitches can also be used to test the robustness of parameter estimation codes in the presence of glitches. Artificially overlapping fake glitches with fake signals will allow us to test if each glitch type produces a bias in the gravitational-wave parameter estimation and to train algorithms which subtract glitches from the data when they occur near real gravitational-wave signals \cite{2022arXiv220513580H, 2018CQGra..35o5017P}.   

\section*{Acknowledgements}
The authors are supported by the Australian Research Council's (ARC) Centre of Excellence for Gravitational Wave Discovery (OzGrav) through project number CE170100004 and through OzGrav Translational Seed Funding. The authors also acknowledge support through ARC LIEF grant LE210100107. JP is supported by Discovery Early Career Research Award (DECRA) project number DE210101050. PDL is supported by ARC Discovery Project DP220101610. This work was performed on the OzSTAR supercomputer which is funded by Swinburne University of Technology and the National Collaborative Research Infrastructure Strategy (NCRIS). We thank the Gravity Spy team for making their data publicly available. This material is based upon work supported by NSF's LIGO Laboratory which is a major facility fully funded by the National Science Foundation.

\appendix

\section{Glitch class balancing}
\label{appendix:balancing}

\begin{figure}
    \includegraphics[width=14cm]{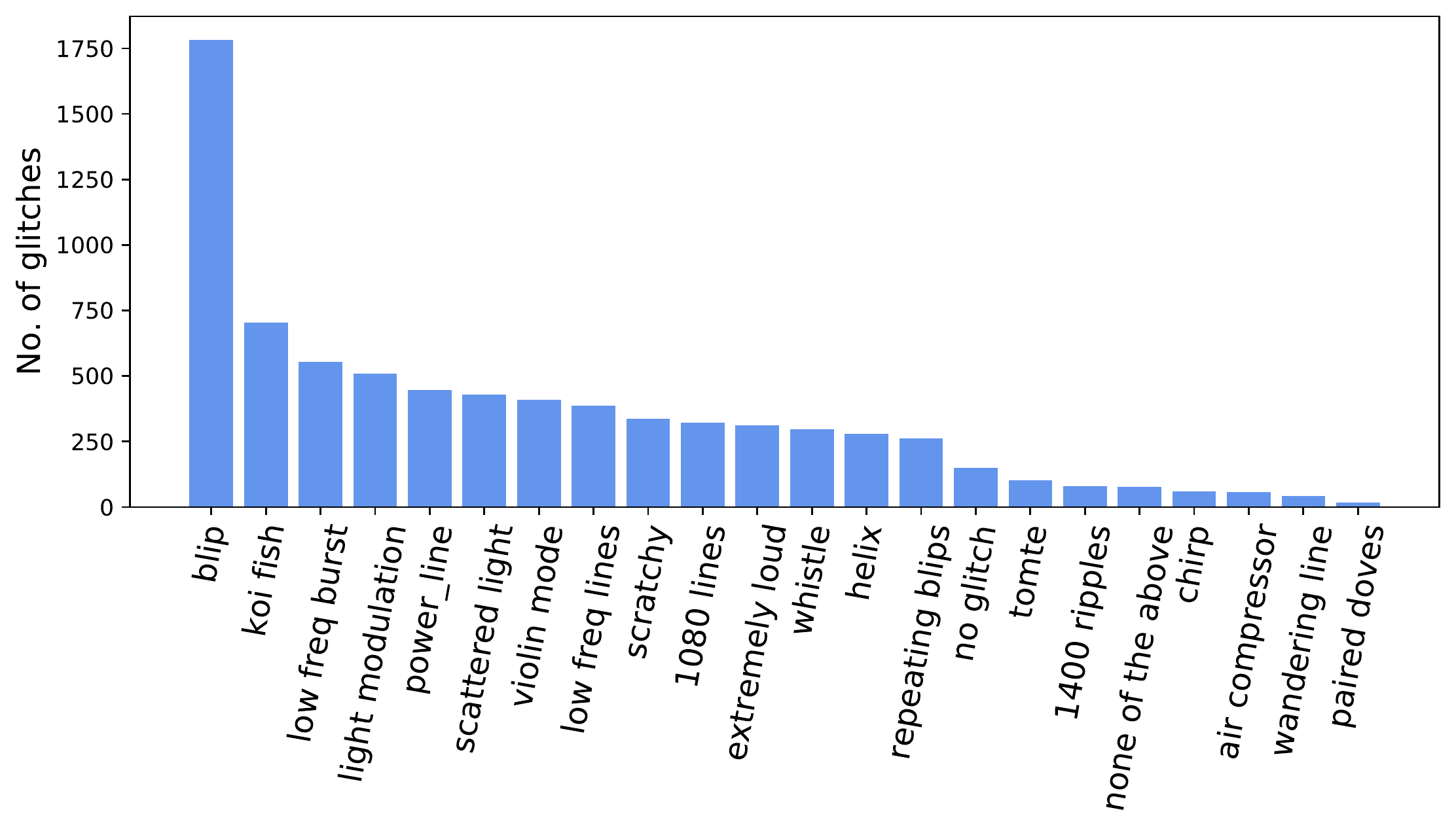}
    \includegraphics[width=14.0cm]{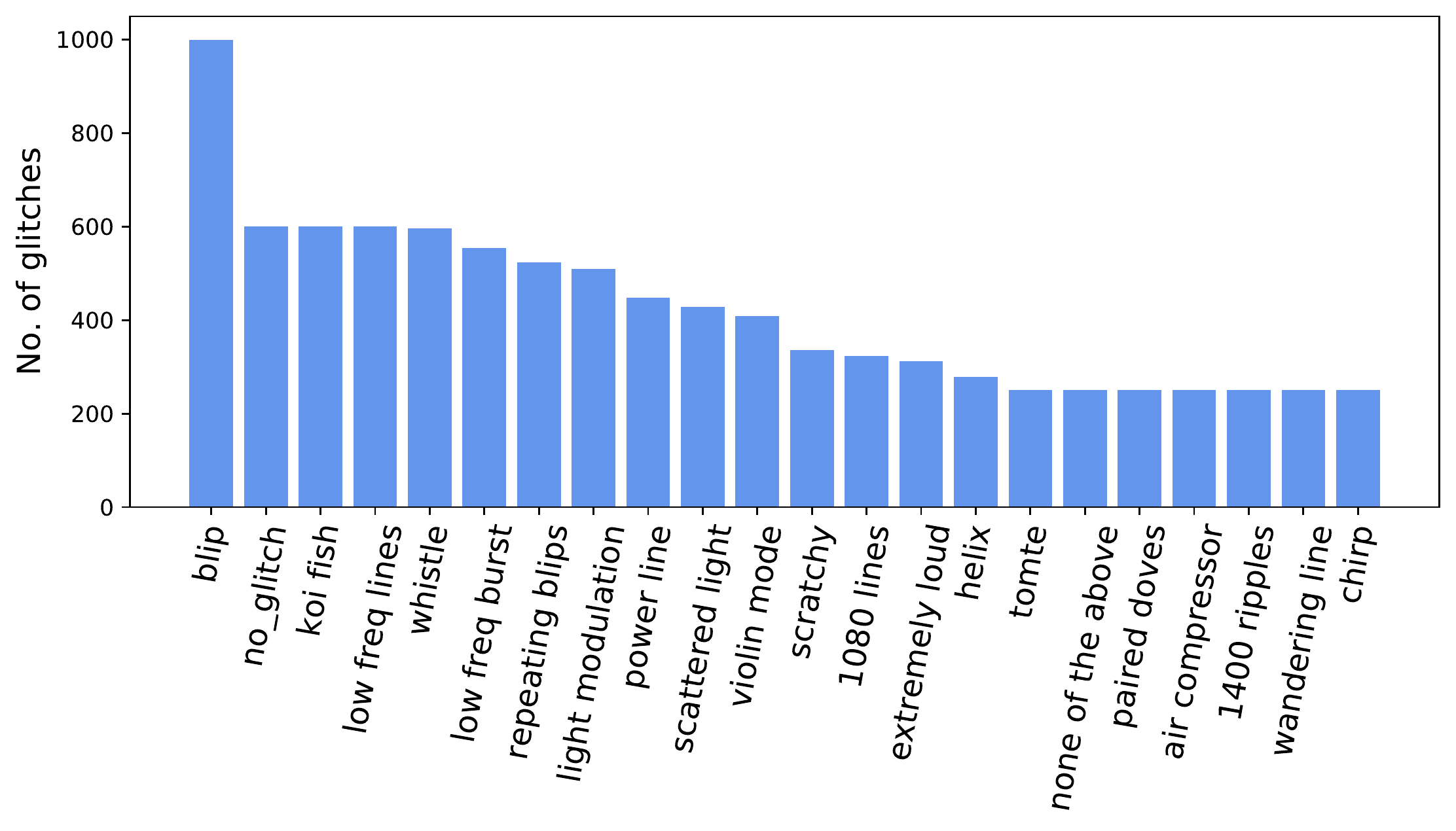}
    \caption{The number of Gravity Spy glitches in each glitch type before (top) and after (bottom) the class balancing. We use the numbers of glitches in the bottom panel to train the CNN.}
    \label{fig:upsample}
\end{figure}

The glitches we use from Gravity Spy have unbalanced classes, which can have a negative affect on machine learning results. Our original class imbalance is shown in the top panel of Figure \ref{fig:upsample}. There are over 1750 glitches in the blip class but only 18 glitches in the smallest glitch category. To prevent this problem from causing classification errors, we change the number of samples for each glitch type before applying the CNN. The new number of glitches is shown in the bottom panel of Figure \ref{fig:upsample}. The blip and koi fish glitches are downsampled and all the other glitches are upsampled, so there is a much more even number of glitches in each category. 

\section{Glitch reproduction}
\label{appendix:all_glitches}

\begin{figure}
    \includegraphics[width=18cm]{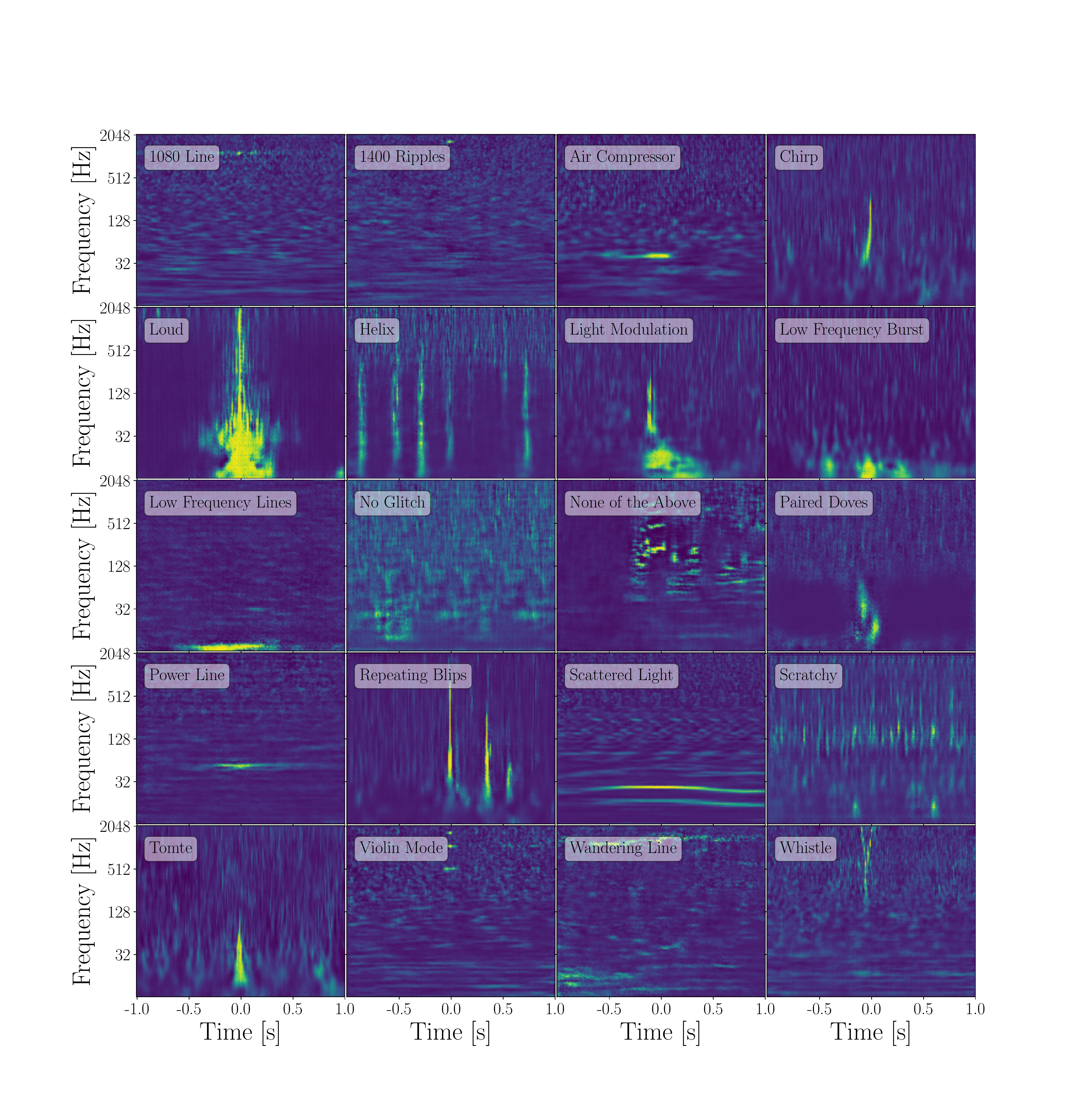}
    \caption{A single GAN-generated image of each type of glitch (excluding Blip and Koi Fish, which are shown in the Figures~\ref{fig:fake_blips} and~\ref{fig:fake_koi}, respectively).}
    \label{fig:all_glitches}
\end{figure}

In Fig.~\ref{fig:all_glitches} we show a single GAN-generated glitch of each class.

\bibliography{bibfile}

\end{document}